\documentclass[preprintnumbers,amsmath,amssymb,superscriptaddress,twocolumn,showpacs]{revtex4-1}
\usepackage{graphicx}
\usepackage{dcolumn}
\usepackage{bm}
\usepackage{physics}
\usepackage[caption=false]{subfig}
\usepackage[english]{babel}
\usepackage{xr-hyper}
\usepackage[hidelinks]{hyperref}

\newcommand{\be}{\begin{equation}}
\newcommand{\ee}{\end{equation}}
\newcommand{\bea}{\begin{eqnarray}}
\newcommand{\eea}{\end{eqnarray}}

\begin{document}

\title{Nuclear Magnetic Resonance with a Levitating Micro-Particle}

\author{J. Voisin$^1$,  A. Durand$^1$, T. Copie$^1$,  M. Perdriat$^1$,  G. H\'etet} 

\affiliation{Laboratoire De Physique de l'\'Ecole Normale Sup\'erieure, \'Ecole Normale Sup\'erieure, PSL Research University, CNRS, Sorbonne Universit\'e, Universit\'e Paris Cit\'e , 24 rue Lhomond, 75231 Paris Cedex 05, France}

\begin{abstract}
Nuclear Magnetic Resonance (NMR) spans diverse fields from biology to quantum science. Employing NMR on a floating object could unveil novel possibilities beyond conventional operational paradigms. Here, we observe Nuclear Magnetic Resonance (NMR) within a levitating micro-diamond using the nuclear spins of nitrogen-14 atoms. By tightly confining the angular degrees of freedom of the diamond in a Paul trap, we achieve efficient hyperfine interaction between optically polarized electronic spins of nitrogen-vacancy centers and the $^{14}$N nuclear spin, enabling nuclear spin polarization and quantum state read-out revealing coherence times up to hundreds of microseconds.
This represents the longest recorded spin coherence time in a levitated system, surpassing previous records by three orders of magnitude. Our results offer promise for various applications, including cooling macroscopic particles to their motional ground state and exploring geometric phases for gyroscopy.
\end{abstract}

\maketitle

Controlling the motion of trapped macroscopic particles in the quantum regime has been the subject of intense research in recent decades. Especially noteworthy is the recent milestone of achieving ground state cooling for a trapped particle \cite{delic2020cooling, tebbenjohanns2021quantum}. However the generation of non Gaussian states required for further quantum control \cite{Aspelmeyer_review}, such as the first phonon Fock state or Schr\"odinger cat states, remains elusive. Employing engineered non-linear trapping potentials have been proposed as a realistic method to generate such states \cite{Roda-Llordes}.  
Another common approach is to transfer the inherent quantum nature of a well controlled two-level system to the mechanical degree of freedom, which can be realized by coupling trapped crystals with embedded spins using magnetic fields \cite{rabl2009strong, Ma2017, horowitz2012electron, delord2017strong, Rusconi2, conangla2018motion, Steiner, Marshman}.
There, the coherence time of the spin system stands as a critical factor, in particular for the preservation of Schr\"odinger cat states \cite{Hou, Tongcang, Marshman}.

A promising approach involves harnessing {\it nuclear spins}, which offers an inherent boost in coherence time over their electronic counterparts \cite{Okazaki, Chen2019, CAO, Steiner}.
However, common techniques for controlling and detecting nuclear spins require sub-Kelvin temperatures, high magnetic fields and large sample volumes, which makes it challenging to realize with a micrometer scale room temperature levitating particle \cite{ANDREW199511, slichter1996}.
A more practical approach is to use materials with embedded optically polarized electron spins, as the hyperfine interaction in these materials allows for the implementation of Dynamic Nuclear Polarization (DNP) protocols \cite{2013PhRvB..88d5306P, Scheuer, Steiner}.  Such techniques present they own practical challenges nevertheless, and remain to be demonstrated in levitation settings.

In this letter, we demonstrate both the polarization and coherent control of nuclear spins of the nitrogen-14 isotope in a levitating diamond using Dynamic Nuclear Polarization (DNP). We measure the nuclear spin coherence time to be in the order of hundreds of microseconds, marking a three-orders-of-magnitude improvement over the longest coherence times previously recorded in levitated systems.
 \begin{figure}[h]
\centering
\includegraphics[width=8.5cm]{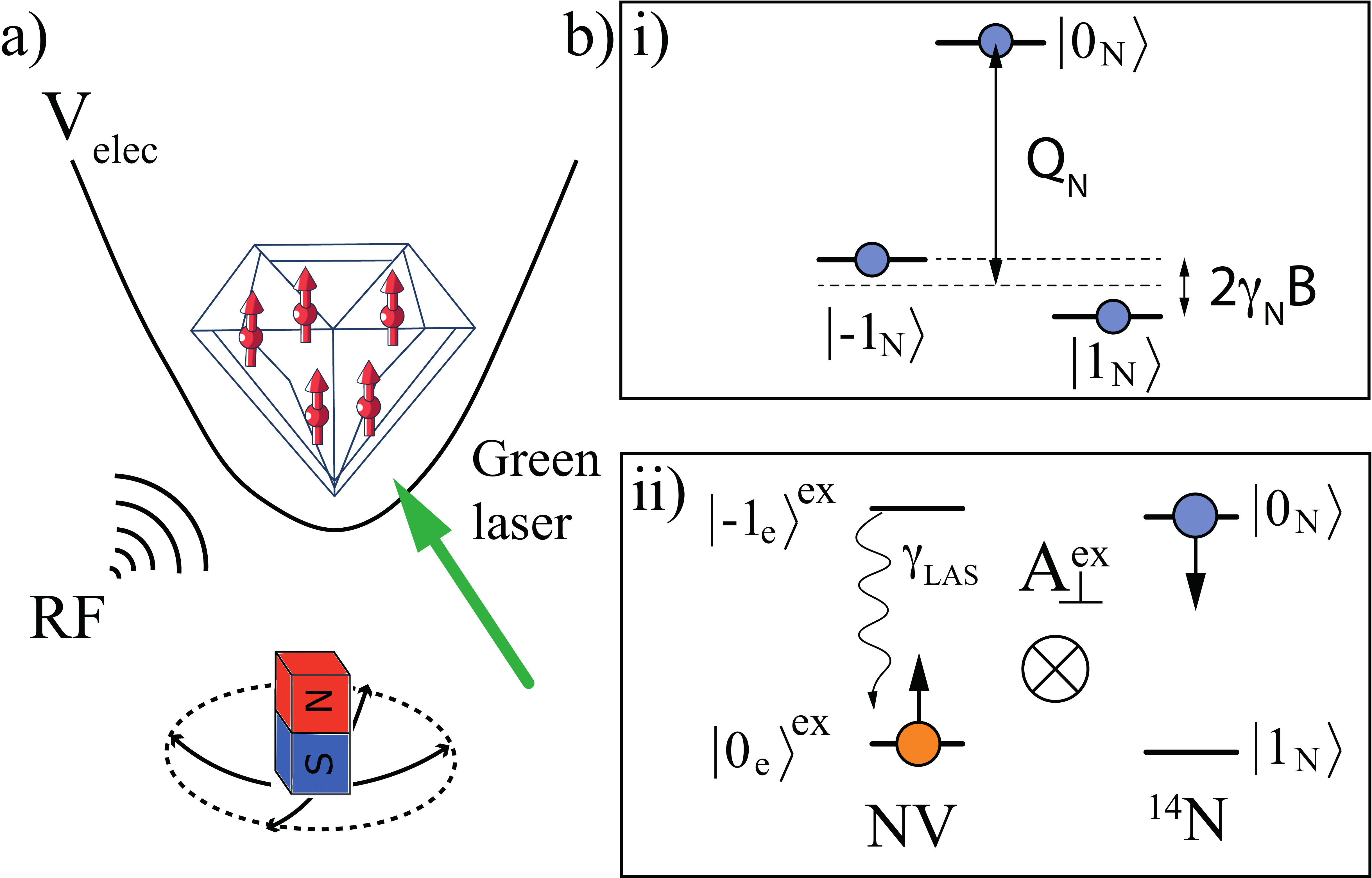}
\caption{a) Sketch of a diamond levitating in an electric potential $V_{\rm elec}$ with embedded nuclear spins (red arrows) subjected to static and radio-frequency (RF) magnetic fields. A green laser is used to polarize the electronic spin of the nitrogen-vacancy centers for Dynamical Nuclear Polarization (DNP). b) i) Level structure of the unpolarized nuclear spin of a $^{14}$N atom, indicating the quadrupolar splitting $Q_N=-4.94$~MHz and the Zeeman splitting between the $\ket{m_I=\pm 1_N}$ spin states. ii)  Flip-flop process (indicated by the two up and down arrows) in the $\{\ket{m_s=0_e}^{\rm ex},\ket{m_s=-1_e}^{\rm ex} \} $ manifold of the optically excited state of the NV center electronic spin, enabling DNP to the $\ket{m_I=1_N}$ state. Wiggly arrow represents fast reset of the electronic spin in the $\ket{m_s=0_e}^{\rm ex}$ state at a rate $\gamma_{\rm las}$, after a flip-flop with the nuclear spin.}
\end{figure}

The employed platform comprises an angularly stable micro-diamond levitating in an electric Paul trap as depicted in Fig.~1-a) (see \cite{delord2017strong, Perdriat2021} for a detailed description of the set-up). 
Here, we use micro-diamonds with 10 to 20 $\mu$m diameter prepared under high-pressure-high-temperature (HPHT). They contain between $10^3$ to $10^5$ nitrogen-vacancy (NV) centers. The negatively charged NV centers
are polarized in the 
$\ket{S=1,m_s=0_e}$ ground state using around 100 ~$\mu$W of green laser light and the photoluminescence (PL) from the embedded NV centers is directed to an avalanche photodiode (APD) (see Sec. I and II of the supplementary materials (SM) \cite{supp}) . 
The PL is used to measure the NV electronic spin magnetic resonances \cite{Delord2017}. Such Optically Detected Magnetic Resonance (ODMR)  is performed by sweeping a microwave signal across one of the NV transitions, thereby bringing the spins to a darker $\ket{S=1,m_s=\pm 1_e}$ state.

Several methods can be employed to polarize nuclear spins in diamond through DNP. 
A very common NMR technique uses co-resonance between nuclear and electronic transitions in the rotating frame by driving them at the same Rabi frequency \cite{slichter1996}. At this so called Hartmann-Hahn condition, cross-polarization is enhanced and nuclear spin polarization is realized. Such a method was successfully applied to polarize the nuclear spins of $^{13}$C through the optically polarized electronic spins of NV centers \cite{Scheuer, Schwartz}.
Other methods use mixing between the NV electronic spins and the $^{14}$N nuclear spins of the NV center by magnetic fields at an angle with respect to the NV axis \cite{HuilleryCPT}. 
These techniques operate efficiently with high quality diamonds prepared by Chemical Vapor Deposition (CVD), where decoherence
is minimized. 
Here, we consider another approach that does not require CVD grown diamonds. The method we use employs cross-relaxation at level crossings tuned by {\it static} magnetic fields \cite{jacques2009}.

We consider DNP of the $^{14}$N atom of the NV center.
The nucleus of $^{14}$N is a spin-1 system with a quadrupole moment $Q_N=-4.94$~MHz aligned to the N-V axis, hereafter labeled as the $z$ axis, with a gyromagnetic ratio $\gamma_N=0.307$~kHz/G. The spin-1 level structure of $^{14}$N is depicted in fig. 1-b)i).
The DNP level crossing occurs in the optically excited state of the NV electronic spin, at the so-called Excited Level Anti-Crossing (ESLAC). The electronic spin states $\ket{0_e}^{\rm ex}$ and $\ket{-1_e}^{\rm ex}$ cross at a magnetic field \(B  \approx 500 \, \text{G}\), applied parallel to the axis of the NV center (see Sec. II of the SM \cite{supp}). 
At the ESLAC, the Hamiltonian describing the hyperfine interaction between the electronic spin $\bm S$ and the nuclear spin $\bm I$ reads 
\bea\label{Eq1}
\hat{H}_{\rm hyp} = \frac{A_\perp^{ex}}{2} (\hat S_+ \hat I_- +\hat S_- \hat I_+)+ A_{zz}^{ex} \hat S_z \hat I_z,
\eea
where $A_\perp^{ex}=A_{xx}^{ex}=A_{yy}^{ex}\approx 20$ MHz are the transverse hyperfine tensor components whereas $A_{zz}^{ex}$ is the longitudinal hyperfine tensor (for more details, see Sec. III in the SM).
The first two terms in Eq. \ref{Eq1} describe electron-nuclei spin flip-flop. In the presence of a green laser, this term allows for effective polarization of nuclear spins through the optical pumping scheme depicted in Fig. 1-b)-ii). Starting from a $\ket{m_s=0_e, m_I=0_N}$ state, the laser brings the electron to the $\ket{m_s=0_e}^{\rm ex}$ state after which a flip-flop changes the electron-nuclear spin state to $\ket{m_s=-1_e^{\rm ex}, m_I=+1_N}$. The NV electronic state is then reset to the $\ket{m_s=0_e}$ ground state while conserving the nuclear spin projection, so after a few optical cycles the system is polarized in the state $\ket{0_e,1_N}$ \cite{jacques2009}.


 \begin{figure}[h]
\centering
\includegraphics[width=8cm]{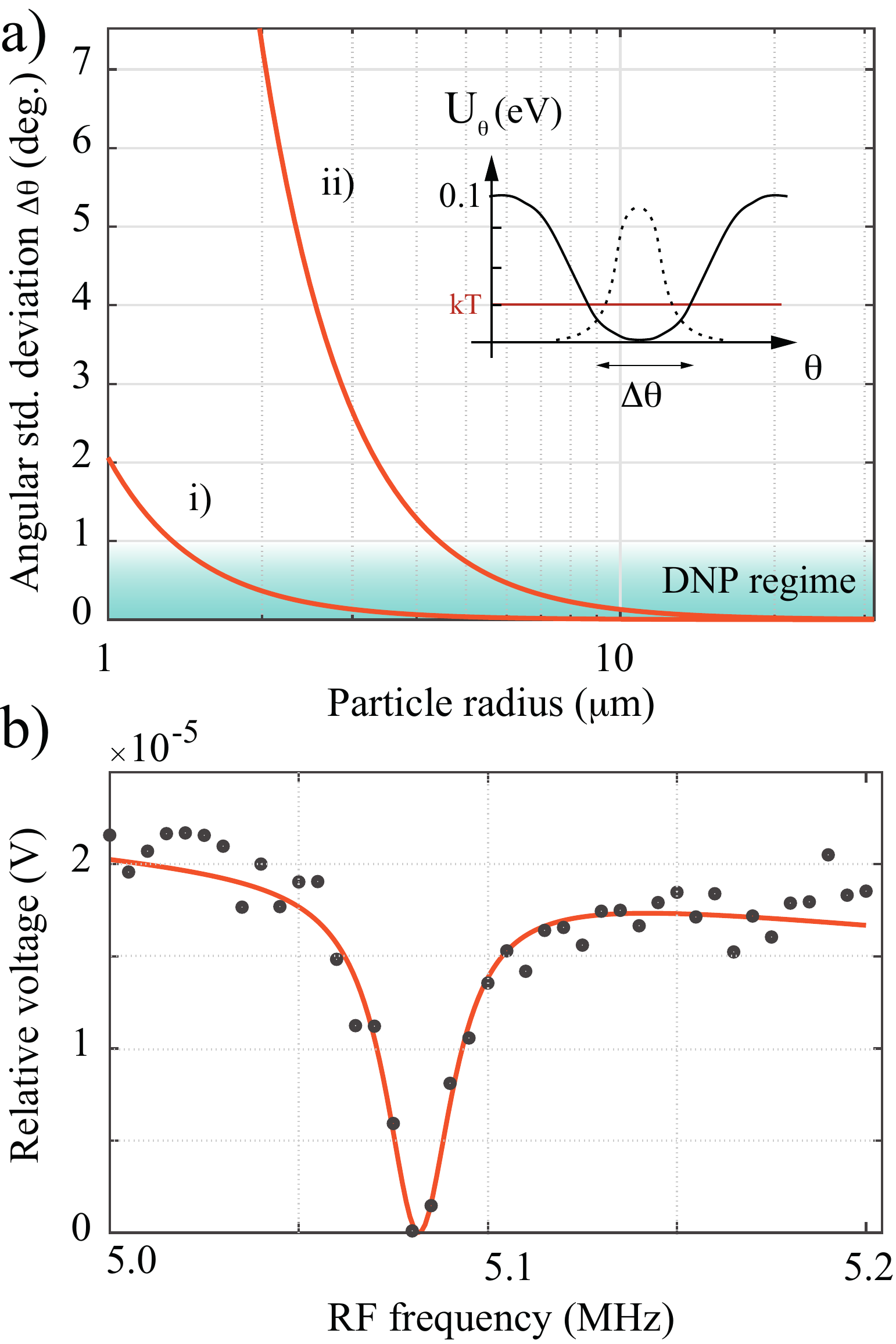}
\caption{a) Standard deviation of the angle of a diamond due to Brownian motion, as a function of its radius. Trace i) and ii) correspond to  $\omega_\theta=(2\pi) 2.0$ kHz and $\omega_\theta=(2\pi) 100$ Hz respectively.   The threshold for efficient DNP ($\Delta \theta \leq 1^\circ)$ is indicated by the blue shaded area where polarization greater than 90\% can be reached \cite{jacques2009}. The figure on the top right shows the angular potential $U_\theta$ of a particle in a Paul trap (plain black line). The dotted line is the Gaussian thermal envelope of the angular degree of freedom at T=300 ~K. b) Black dots show data of optically detected nuclear spin polarization in a trapped micro-diamond. Straight line is a Lorentzian fit including a linear rescaling (see text and Sec. VI in the SM for more details \cite{supp}). The trace was offset by 0.068~V for clarity. }
\label{conf}
\end{figure}

Levitation imposes a significant constraint compared to using DNP with fixed diamonds, primarily due to the Brownian motion of the angular degrees of freedom. It was shown that the three angles of a diamond particle with non-zero quadrupolar moment can be confined in asymmetrical Paul traps \cite{delord2017strong}.
The variance of an angle $\theta$ reads $\Delta \theta^2=kT /I \omega_\theta^2$, where $kT$ is the gas thermal energy, at thermal equilibrium with the particle motion, $I$ the particle moment of inertia and $\omega_\theta$ is the eigen-frequency \cite{delord2017strong}. Fig. \ref{conf}-a) shows the expected standard deviation $\Delta \theta$ as a function of the diamond radius. 
 Trace i) and ii) are derived by taking the upper and lower bounds for angular frequencies $\omega_\theta=(2\pi)2.0$~kHz and $\omega_\theta=(2\pi)100$~Hz respectively that are typically found with librating diamonds in micro-Paul traps \cite{delord2020spincooling}. 
For a given Paul trap potential, the difference between the particle angular frequencies is determined by the charge density on the surface of the particle (between $10^2$ to $10^4$ elementary charges), as well on as the particle shape \cite{delord2017strong}.  

Angular motion makes polarization at the ESLAC challenging because it requires angular deviations between the N-V and magnetic fields axes to be less than one degree \cite{jacques2009}.
Fig.~\ref{conf}-a) shows that to satisfy $\Delta \theta \lesssim 1^\circ$, one must choose particle diameters $d \gtrsim 8~\mu$m.
Furthermore, it was observed that extraneous optical and electric torques also lead to angular motion \cite{Delord2016} that prohibit efficient DNP. After successful loading of a particle with the appropriate diameter, the laser field, the RF Paul trap drive voltage and the RF frequencies were therefore tuned below the rotation instability thresholds \cite{2023arXiv230901545P}.

In addition to angular motion, another major difficulty of NMR with trapped diamonds is that the orientations of the electronic and nuclear spin axes relative to the laboratory frame are unknown prior to particle loading. After each loading run, particles have a different shape, implying different surface charge distributions and thus different orientations of the crystalline axes with respect to the main trap axis. To align the magnetic field along one of the four NV axes,  a two-axis goniometer that holds a NdFeB permanent magnet with a one-inch diameter was designed.
The goniometer allows for tuning the angle of the magnet with a solid angle corresponding to  $\approx 45^\circ$ spans for two perpendicular rotation axes (see Fig. 1-a)) with an angular resolution below 1 degree. In addition, the magnet is mounted on a translation stage attached to one of the rotation stage so that the translation axis intersects the trap center. This setup allows for tuning the magnetic field strength once the angle is set close to an NV axis. A dedicated chamber was also built to enable the manipulation of the goniometer and translation stage without imparting air currents to the particle. This precaution prevents particle loss or changes in the particle angle during magnetic field alignment.

The alignment procedure consists in : i) setting the B field to around 100 G at the diamond location.
This is achieved simply by tuning the magnet distance to about 10 cm from the trapped diamond.
ii) Aligning the magnet angle to one of the NV axes, {\it i.e.} along one of the $\langle 111 \rangle$ diamond crystalline directions, 
while recording ODMR spectra.  When the magnetic field axis is close to a $\langle 111 \rangle$ direction, ODMR indicates that the three classes that form an angle of 109 degrees with respect to the targeted NV axis are degenerate.
This procedure, detailed in the section IV the SM \cite{supp}, must be implemented at modest magnetic fields to minimize state mixing by the transverse component of the magnetic field. Such mixing would otherwise reduce the ODMR contrast of the three auxiliary classes. Then, iii), the magnet is translated towards the trapped diamond to reach the ESLAC. During this process however, the NV and magnetic field axes become slightly misaligned. In magnetic fields $\abs{\bm B}> 300$~G, the ODMR contrasts of the auxiliary NV centers become very weak, making it impractical to use the technique from step ii) for correcting misalignment.
To retain alignment close to the ESLAC, we resort to a cross-relaxation between NV centers and the electronic spin of $P_1$ centers (substitutional nitrogen atoms). This mechanism leads to a substantial reduction of the NV PL at a NV-P$_1$ transition degeneracy where $\abs{\bm B}\approx$ 500 G \cite{Armstrong}. There, the PL drop is very sensitive to magnetic field mis-alignment $\delta \theta$ with respect to the NV axis, giving rise to a Lorentzian dip as a function of $\abs{\bm B}$ with a width that depends linearly on $\delta \theta$ \cite{epstein2005anisotropic}. The step iv) thus consists in maximizing the NV PL when the magnetic field is a few Gauss smaller than the  NV-P$_1$ co-resonance, providing sub-degree precision between the NV and magnetic field axes in a few seconds.
When successful, at the end of these four sequential operations, nuclear spins are polarized in the $\ket{m_I=+1_N}$ state. Although the optimum point for resonant cross-polarization in CVD-grown diamonds is approximately 510 G, in HPHT-grown samples the aforementioned cross-relaxation between the polarized NV centers and the unpolarized $P_1$ centers reduces the NV relaxation time ($T_1$). This effect in turn drastically diminishes the nuclear spin polarization \cite{Jarmola_1}, so the experiment was conducted at a slightly lower magnetic field $\abs{\bm B}\approx 450$~G.

Once the magnetic field is aligned, we leverage the dependence of NV center photoluminescence on nuclear spin states to measure spin polarization. This approach underpins the technique known as Optically Detected Nuclear Magnetic Resonance (ODNMR) \cite{Jarmola_1}. In continuous wave ODNMR, the laser is continuously applied and thus always brings a fraction of the electronic spin population to the excited state 
$\ket{m_s=0_e}^{\rm ex}$ in which, at the ESLAC condition, electron-nuclear spin flip-flops occur, as {\it per} Eq. \ref{Eq1}. After flipping to the $\ket{m_s=-1_e}^{\rm ex}$ state, the electron will non-radiatively decay to the NV metastable state, thereby lowering the PL rate. Overall, when an RF tone is  resonant with the $\ket{m_s=0_e, m_I=1_N}$ to $\ket{m_s=0_e,m_I=0_N}$ nuclear spin transition, the PL is indeed expected to drop. 

\begin{figure}
\includegraphics[width=9cm]{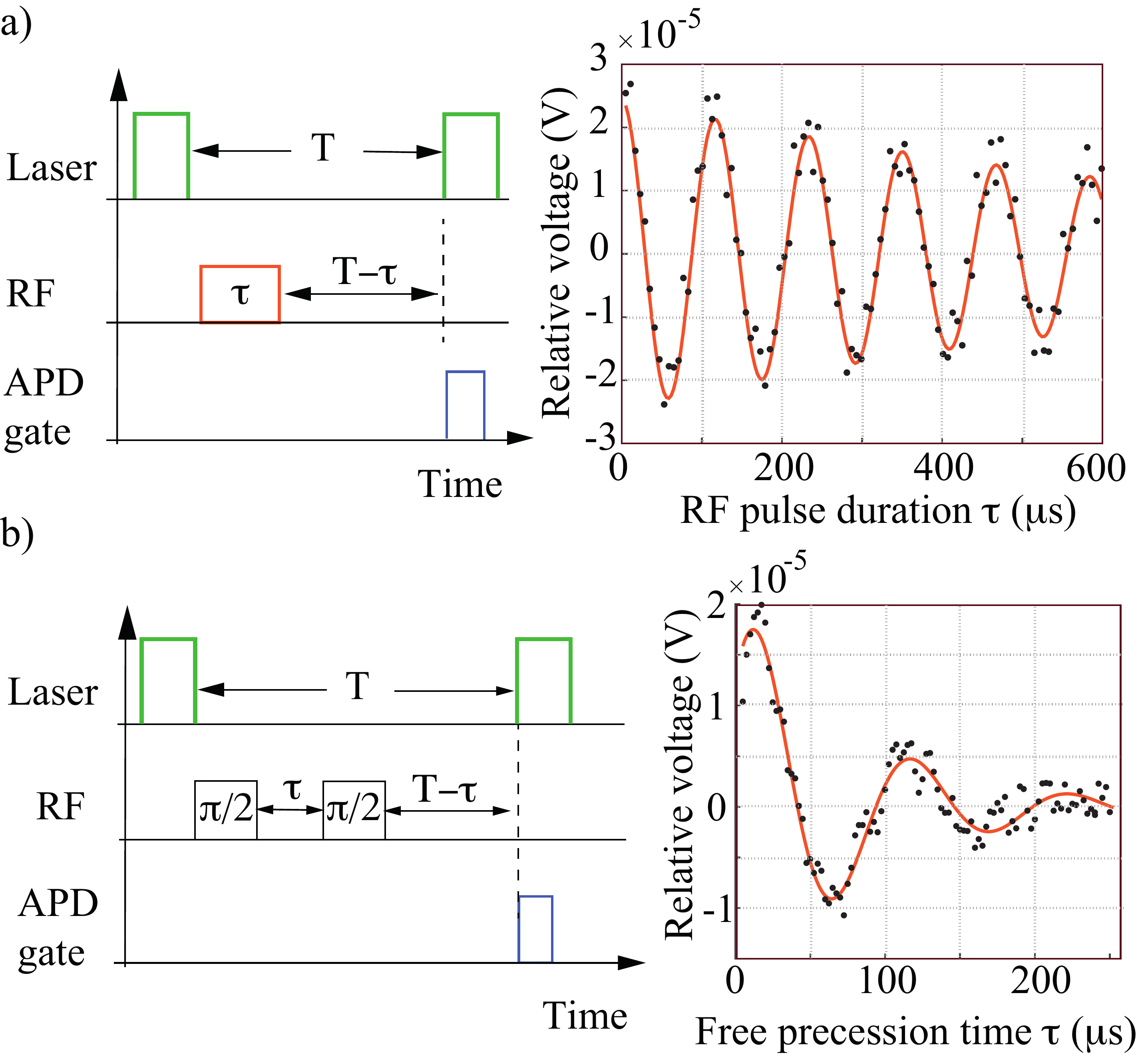}
\caption{a) Rabi and b) Ramsey oscillations of the nuclear spin of $^{14}$N atoms in a trapped diamond, detected using the photoluminescence from the nitrogen-vacancy centers electronic spins. The sequences that are employed in both experiments are shown on the left panels. To avoid artifacts, an extra delay $T-\tau$ between the RF signal and the detection pulses is used to maintain a fixed duration $T$ in the dark (see Sec. VII of the SM \cite{supp}).}
\label{rabi}
\end{figure}

Fig. \ref{conf}-b) shows the NV PL as a function of the RF drive across the $\ket{m_I=1_N}\leftrightarrow \ket{m_I=0_N}$ nitrogen nuclear spin transition   in a trapped diamond. 
A clear PL dip is observed at 5.08 MHz, consistent with $|Q_N|+\gamma_N B\approx 5.08$~MHz, indicating nuclear spin driving. Similar ODNMR scans were observed when using other diamond particles as well.
The asymmetry in the NMR signal is also present on NMR scans performed on static particles (see Fig. 3 in Sec. VI. of the SM) and may be attributed to the interaction between the two nuclear spin transitions in the $\Lambda$ scheme of the $\ket{m_s=0_e}$ manifold. Both transitions are indeed separated by a few 100 kHz only, and reach the same $\ket{m_I=0_N}$ level.

Ideally, NMR contrasts can reach up to 2$\%$ in NV ensembles. The measured $30$ m\% contrast is partly related to the above-mentioned choice of the B field which is tuned slightly off the ESLAC. The  nuclear spin polarization is then reduced compared to when using CVD grown samples \cite{Jarmola_1}, where the concentration of P$_1$ centers is two orders of magnitude lower. Another reason is the relatively low green laser intensity that is used to polarize and read out the nuclear spin states compared to state of the art ODNMR. A too high laser intensity ($>$ 1W/cm$^2$) was seen to result in a significant photoionisation of the particle surface charges in our experiment. This leads to a change in the particle angle on time scales of several minutes, preventing efficient signal averaging at a given magnetic field angle with respect to the NV axis. 

To estimate the coherence time $T_{2,N}^*$ of the nuclear spins, we use pulsed measurements where the laser excitation is separated temporally from the RF excitation \cite{DreauAvoiding}.
 Fig. \ref{rabi}-a), right panel, shows Rabi oscillations of the nuclear spins, measured using the sequence that is depicted on the left panel. The nuclear spins are polarized using $\approx 10\mu$W of laser light, after which a radio-frequency field resonant with the nuclear spin transition is applied. The laser is then turned on again to detect the nuclear spin population using the NV photoluminescence. The sequence is repeated with increasing RF pulse duration $\tau$ ranging from 0 to 600 $\mu$s (see sec. VII of the SM \cite{supp} for an analysis of the Rabi decay time).
Fig.~\ref{rabi}-b) shows a Ramsey measurement as well as the employed sequence (left panel). The measurement was conducted with two RF pulses detuned by 10~kHz from the nuclear spin resonance to enable spin precession about the NV axis. The NV PL is then measured for increasing time intervals between the two RF $\pi/2$ pulses.
A fit to the data reveals a decay time $T_{2,N}^*\approx 120~\mu$s.  Importantly, the longest reported coherence times for spins in a trapped crystal are within $ 60$ to $100~$ns using the electronic spins of nitrogen-vacancy centers in a diamond \cite{delord2018ramsey, Jin2024, Pettit}.  The present result demonstrates more than three orders of magnitude longer $T_2^*$ \footnote{$T_{2,e}^*/T_{2,N}^*=7.5\times 10^3$ is very close to the ratio between the gyromagnetic ratios of the electronic and nuclear spins in similar diamond crystal growth techniques. This indicates that the source of decoherence for both spin species is likely to be the fluctuating $P_1$ spin bath.}, which could be improved further using CVD grown samples where the $P_1$ concentration is drastically reduced.

Our experiment opens a path towards spin-mechanical experiments with trapped particles in regimes that were previously not accessible.  
In the reference frame of an RF field driving a two-level spin system, the two levels are split by the Rabi frequency $\Omega_{\rm RF}$.
When the spin system is coupled to the motion of the particle in a harmonic potential with a center of mass frequency $\omega/2\pi$, resonant spin-mechanical driving occurs when 
$\Omega_{RF} \approx \omega$ \cite{Tongcang, rabl2009strong, delord2017strong, Ge}. This regime is currently untouched in the field of levitodynamics because of the typically large spin decoherence rate compared to the mechanical frequency. In our experiment, we measured a Rabi frequency $\Omega_{\rm RF}= (2\pi)8.5~$kHz, 
which is on the same order of magnitude as the diamond center of mass modes, offering prospects for high-fidelity quantum non-demolition spin readout \cite{Didier}. 
In this regime, spin cooling to the motional ground state {\it via} red detuned micro-wave tone can also be envisioned, in strong analogy with the sideband cooling cooling of trapped ions and opto-mechanical systems \cite{Aspelmeyer_review}.  The mechanical frequencies of the diamonds in the Paul trap are currently in the kHz range. Increasing this figure by an order of magnitude through higher charge densities and/or even smaller traps will enable the spin mechanical system to be operate in the sideband resolved regime. 

Beyond applications in spin-mechanics, let us add that embedded spin qubits are capable of detecting the pseudo-magnetic field related to the Barnett effect and the quantum geometric phase associated with particle rotation \cite{chudo2014observation, Wood2020, Maclaurin2012, Chen2019}. This spin-mechanical rotation coupling can be leveraged using nuclear spins to develop highly sensitive gyroscopes and rotational matter-wave interferometers \cite{Zhang}.
In addition, the nuclear spin population lifetime being in the second range, rotating the diamond \cite{2023arXiv230901545P} whilst nuclear spins are polarized may open a path towards efficient magic angle spinning \cite{ANDREW_MAS, ANDREW199511, Marti}, boosting further the nuclear spin coherence time.\\

\begin{acknowledgments}
We thank Pascal Morfin for technical support and Hendrick Ulbricht for stimulating discussions. 
M.P.\ and G.H.\ have been supported by Region Île-de-France in the framework of the DIM SIRTEQ.
This project was funded within the QuantERA II Programme that has received funding from the European Union’s Horizon 2020 research and innovation programme under Grant Agreement No 101017733.
\end{acknowledgments}

\bibliography{NMR.bib}

\end{document}


\title{Nuclear Magnetic Resonance with a Levitating Micro-Particle \\
--- Supplementary Materials --- }

\author{J. Voisin$^1$,  A. Durand$^1$, T. Copie$^1$,  M. Perdriat$^1$,  G. H\'etet} 
\affiliation{Laboratoire De Physique de l'\'Ecole Normale Sup\'erieure, \'Ecole Normale Sup\'erieure, PSL Research University, CNRS, Sorbonne Universit\'e, Universit\'e Paris Cit\'e , 24 rue Lhomond, 75231 Paris Cedex 05, France}

\maketitle

In this supplementary materials we describe the electronic spin structure of the NV center, the principle of dynamical nuclear polarisation (DNP) and the method that was deployed to align the magnetic field in the DNP levitation experiment. 

\section{NV$^-$ Electronic structure}

The unit cell of diamond with a nitrogen vacancy center is depicted in the inset of Fig. 1-i).

Both the ground and optically excited electronic states of the NV$^-$ center are triplet states with zero-field splittings that set a natural quantization axis $z$ along the N-V direction \cite{DOHERTY20131}. 
The spin-orbit interaction is averaged at room temperature in the excited state, through a dynamic Jahn-Teller process coupled with the phonons of the diamond matrix.  The spin structure in the excited state of the NV center is thus analogous to that of the ground state (see Fig. 2). 
In the absence of magnetic field, the $\vert S=1, m_s=\pm 1_e \rangle$ states both lie $D\approx(2\pi) 2.87$ GHz and $D_\text{ex}\approx(2\pi) 1.40$ GHz above the $\vert m_s=0_e \rangle$ state in the ground and excited states respectively, at room temperature.  Importantly, the NV center electronic spin can be optically polarized to the $\ket{S=1, m_s=0_e }$ state using green laser light.
Further, the PL is larger in the $\ket{m_s=0_e}$ state than in the $\ket{m_s=\pm 1_e}$ states so that applying a microwave signal at this frequency results in a drop of the photoluminescence (PL). This is the essence of Optically detected magnetic resonance 
(ODMR).

\section{ODMR measurements}

In this section we describe the electronic spin state detection technique in detail. In this experiment, we use a confocal spectroscopy setup similar to the one used in ref.~\cite{Perdriat2021}, that allows green laser excitation and red photoluminescence collection using only one lens.

To detect electronic and nuclear spin state transitions, we use continuous wave optically detected magnetic resonance (ODMR). Microwave and RF excitation is provided by a Rohde and Schwarz SMB100A signal generator. The RF signal for nuclear spin state detection is then amplified using a Mini-Circuit LZY-22+ amplifier. 

\begin{figure}[h]
\centering
\includegraphics[width=12cm]{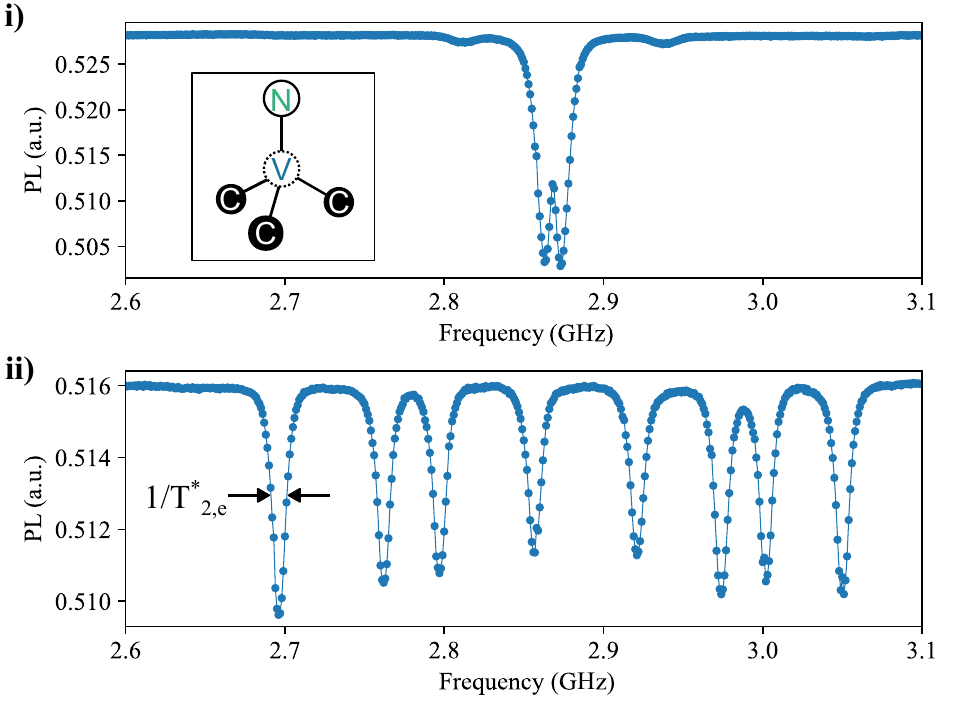}
\caption{Photoluminescence from NV centers in a diamond as the function of a microwave frequency, without i) and with ii) an arbitrary magnetic field.}\label{ODMR_clean}
\end{figure}

Fig. \ref{ODMR_clean}-i) shows an ODMR realized on a diamond with no magnetic field applied. At zero magnetic field, the $\ket{m_S=+1}$ and $\ket{m_S=-1}$ states are degenerate. However, even at zero field, there is a small splitting due to strain in the microdiamond. The two asymmetrical sidebands are due to hyperfine interactions between the electronic spin of the NV center and the carbon 13 atoms in the first adjacent cell. The inset shows one of the possible N-V orientations in the crystal lattice. There are in fact three other possible orientations for the N-V axis in the diamond. The nitrogen atom can be at any carbon position adjacent to the vacancy in the figure. These four positions are called classes and give rise to four different projections of the magnetic field along the NV axis. When a magnetic field is applied, the degeneracy between the four classes is lifted and one typically observes 8 different spin transitions instead of two, as shown in Fig. \ref{ODMR_clean}-ii).

 \begin{figure}[h]
\centering
\includegraphics[width=12cm]{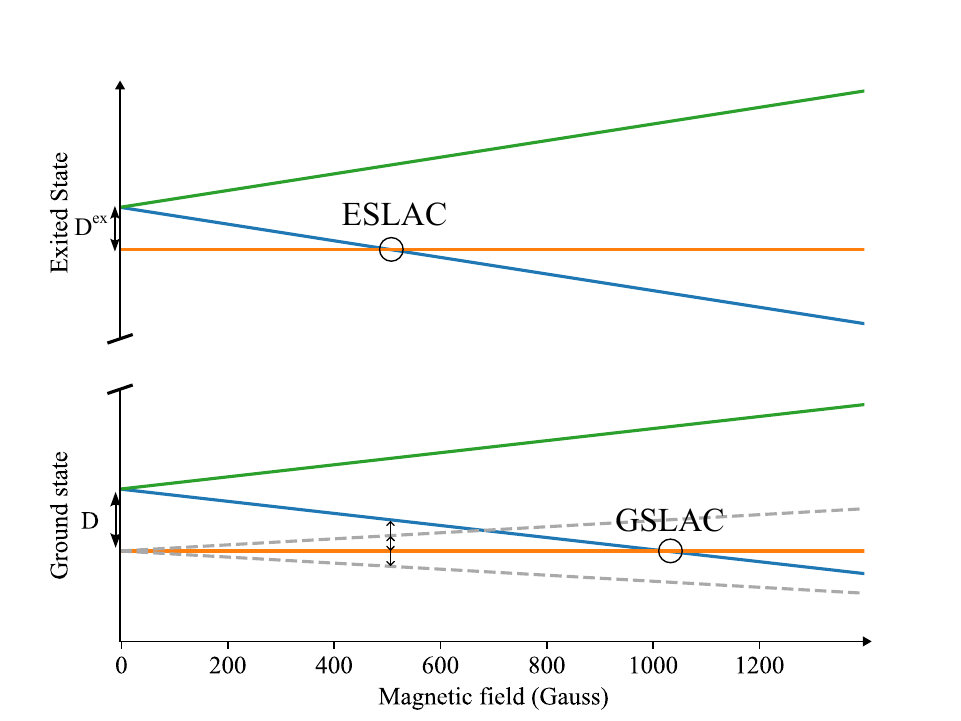}
\caption{Electronic spin state energy levels as a function of the magnetic field. The green, blue and orange lines correspond to the energy levels of the $\ket{+1_e}$, $\ket{-1_e}$ and $\ket{0_e}$ states respectively. The dashed grey lines shows the energy evolution of the $P_1$ electronic spin states $\ket{+1/2}$ and $\ket{-1/2}$. Black arrows at 510 Gauss indicate energy matching between $P_1$ electronic transitions and the $\ket{m_s=0_e}$ and $\ket{m_s=-1_e}$ transition in the NV ground state. $D$ and $D_\text{ex}$ are the zero field splittings in the ground and excited states respectively.}\label{ESLAC}
\end{figure}



Neglecting strains in the diamond matrix and considering a static magnetic field aligned parallel to the NV center axis, denoted along the $z$ direction, the Hamiltonian of the system in the excited state is written as
\[
\hat{H}_{0, \text{ex}} = D_{\text{ex}}(\hat{S}_z)^2 + \gamma_{{e}}B\hat{S}_z 
\]
where \(D_{\text{ex}} \approx 1.43 \, \text{GHz}\) is the spin-spin interaction constant in the excited state. Note that the gyromagnetic ratio is identical in the ground and excited states, as the orbital angular momentum does not intervene to first order in the system dynamics (orbital momentum quenching). 

Fig. \ref{ESLAC} shows the energy levels of the electronic spin states of the NV center, both in the optically excited state and in the ground state. 
The electronic spin states 
 cross for a magnetic field \(B = D_{\text{ex}}/\gamma_{{e}} \approx 500 \, \text{G}\), applied parallel to the axis of the NV center. This level crossing in the excited state (the ESLAC) allows for effective polarization of nuclear spins interacting with the NV center through optical pumping as we shall now describe.

\section{Principle of Dynamical Nuclear Polarisation of $^{14}$N}

The nonspherical symmetry of the nucleus charges in the crystal field of the NV center together with the spin 1 character of the $^{14}$N nuclear spins imply that they align to the NV axis.
Let us first consider the total hamiltonian of the coupled nuclear and electronic spin in its $\ket{S=1, m_s=0_e}$ state. The electronic spin hamiltonian is described by $\hat{H}_0$.
Assuming that the magnetic field is along the NV axis, the Hamiltonian \(\hat{H}\) of the total system can be written as:
$$
\hat{H} = \hat{H}_0 + \gamma_N  B   \hat{I}_z + Q_N (\hat I_z)^2 $$ where \(\hat{I}\) is the operator associated with the nuclear spin of nitrogen. The second term of the Hamiltonian is the Zeeman interaction term associated with the nuclear spin of nitrogen, where $\gamma_N \approx - 0.307 \, \text{kHz} \cdot \text{G}^{-1}$ is the gyromagnetic ratio of the \(^{14}N\) nuclear spin. The third term is an electric quadrupole interaction term. The electric quadrupole interaction constant is \(Q_N \approx -4.94 \, \text{MHz}\).  The hyperfine interaction in the NV ground state $\ket{S=1, m_s=0}$ was neglected here, which is a good approximation away from the ground state level crossing \cite{Sangtawesin}.\\

We will now consider the hyperfine interaction between the nuclear spin of the $^{14}$N nucleus with the electronic spin of the NV center when brought to the optically excited state.
The main reason for considering the ESLAC rather than the GSLAC is that the electronic density is closer to the nucleus in the electron excited state, boosting the hyperfine interaction by an order of magnitude. Further, the ESLAC can be reached with a smaller magnetic field. 



The spin dynamics under optical excitation leads to polarization of the electronic spin of the NV center in the $\ket{0_e}$ state. Since transitions to the metastable state are not sensitive to the projection of the nuclear spin states, they are not {\it a priori} affected by this dynamics. However, the presence of an interaction mixing electronic and nuclear spin states can enable polarization of the nuclear spins through optical pumping of the NV center's electronic spin. The perpendicular hyperfine coupling, enhanced during a crossing of electronic spin levels in the excited state of the NV center, can produce this mixing and allows for polarization of nuclear spins under optical excitation as well as a drop of PL when the nuclear spin state flips.

We now add the terms related to the hyperfine interaction with a nuclear spin. The Hamiltonian of the coupled system is then written as:
\[
\hat{H}_{\text{ex}} = \hat{H}_{0,\text{ex}} + \gamma_N B\hat{I}_z + \hat{S}\cdot\mathbf{A}^{\mathbf{\text{ex}}}\cdot\hat{I}
\]
where \(\mathbf{A}_{\text{ex}}\) is the hyperfine tensor of the interaction between the electronic spin and the nuclear spin in the excited state. 

This interaction includes both the Fermi contact interaction and the anisotropic magnetic dipole interaction. In the $x,y,z$ basis attached to the NV center, the hyperfine tensor $\underline{\underline{\mathbf{A}}}^\mathbf{\text{ex}}$
is diagonal because nitrogen possesses the same symmetry as the NV center. 
 The Hamiltonian describing the hyperfine interaction then expands to 
 \[
\hat{H}_{\rm hyp} = \frac{A_\perp^\text{ex}}{2} (\hat S_+ \hat I_- +\hat S_- \hat I_+)+ A_{zz}^\text{ex} \hat S_z \hat I_z
\]
where $A_\perp^\text{ex}=A_{xx}^\text{ex}=A_{yy}^\text{ex}$.


We can restrict our analysis to the electronic spin multiplets $\ket{0_e}^{\text{ex}}$ and $\ket{-1_e}^{\text{ex}}$ because the $\ket{+1_e}^{\text{ex}}$ state is very far above the other states around the level crossing in the excited state (see Fig. \ref{ESLAC}). 
%
%

Let us describe the nuclear polarization mechanism. Optical pumping allows for the polarization of the electron spin of the NV center into the state $\ket{0_e}$, through non-radiative transitions to a metastable singlet state. We are therefore only interested in optical transitions starting from the states $\ket{0_e,+1_N}$ and $\ket{0_e,0_N}$ of the ground level. The nuclear spin is initially in a statistical mixture of the three states $\ket{+1_N}$, $\ket{-1_N}$  and $\ket{0_N}$. When the system is far from the level crossing in the excited state, optical transitions perfectly conserve the projection of the nuclear spin state, since no spin state mixing occurs in the excited state. The nuclear spin cannot be polarized, and the system remains in an equiprobable statistical mixture of the three states $\ket{0_e,+1_N}$, $\ket{0_e,-1_N}$ and $\ket{0_e,0_N}$.

This situation is drastically altered near the level crossing, when a magnetic field of magnitude $B \sim 510 \, \text{G}$ is applied along the NV axis.
Indeed, when the system is pumped optically into the excited state $\ket{0_e^{\text{ex}},0_N}$, a joint flipping of the electron spin and the nuclear spin, $\ket{0_e^{\text{ex}},0_N} \rightarrow \ket{-1_e^{\text{ex}},+1_N}$, can be induced by the perpendicular component $A^{\text{ex}}$ of the hyperfine tensor. This nuclear spin flipping can then be transferred to the ground state by the electron spin state polarization process. The optical transition from the state $\ket{0_e,+1_N}$, on the other hand, always conserves the nuclear spin projection. If the system starts from the $\ket{0_e^{\text{ex}},-1_N}$, it will end up in the $\ket{0_e,0_N}$ spin state of the ground state, where it will follow the same steps as depicted before. Thus the mixing of excited states of electron and nuclear spins due to the perpendicular hyperfine component efficiently polarizes the system in the ground state $\ket{0_e,+1_N}$ after a few optical cycles.
The mechanism is sketched in the fig. 1.a)-ii) of the main text. 

Under a B field of 510 G and relative angles between NV and magnetic axes below 1 degree, the nuclear spins of nitrogen atoms in diamond can then be polarized in the $\ket{m_I}=\ket{+1_N}$ state through their hyperfine coupling to the NV electronic spins, at room-temperature \cite{jacques2009}. 

\section{Alignment Procedure}

Aligning the magnetic field is the primary challenge in this experiment. This section provides a detailed description of the alignment procedure.

As outlined in the main text, the alignment is executed in four steps. We utilize the QuDi software suite \cite{BINDER201785}, which allows for real-time recording of ODMR spectra. At magnetic fields in the range of hundreds of Gauss, we tune the magnetic field and monitor the transition frequencies of the four NV center classes.

\begin{figure}[h]
\centering
\includegraphics[width=12cm]{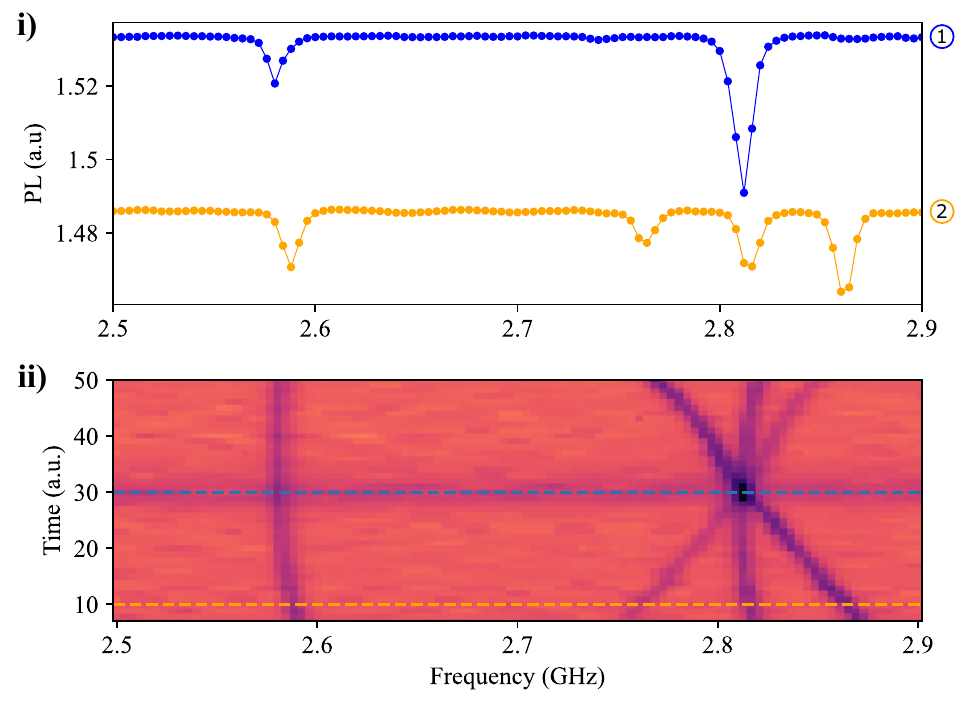}
\caption{i) Photoluminescence from NV centers as a function of the microwave frequency. The blue curve (1) shows an ODMR with a magnetic field of 100 Gauss aligned to one of the four NV classes. The orange curve (2) corresponds to an ODMR with a slightly misaligned magnetic field. ii) 2D map of successive ODMRs over time where the B field angle is tuned slightly away from the $\langle 111 \rangle$ axis. The blue, resp orange, dashed horizontal lines correspond to the situation 1 (resp. 2) in the above ODMRs.}\label{alignment_ODMR}
\end{figure}

Figure \ref{alignment_ODMR}-i) shows the photoluminescence of NV centers as a function of the microwave frequency for two different magnetic field alignments. The blue curve shows an ODMR where the magnetic field is aligned with one of the four NV axes, while the orange curve is when a magnetic field misaligned by a few degrees. By observing the electronic spin state transition frequencies (dips in the photoluminescence), we can estimate the alignment of the NV axis with the magnetic field. Figure \ref{alignment_ODMR}-ii) displays a 2D map of successive ODMR spectra over time as a function of the microwave frequency. During acquisition, the magnetic field is rotated by a few degrees, causing shifts in the spin state transition frequencies. The optimal alignment is indicated at 30 (a.u.), when the three classes are degenerate. This method serves as the initial alignment tool for our experiment.

Once the three NV center classes are degenerate, we employ a secondary alignment method based on the cross-relaxation properties of NV centers with $P_1$ centers. As the magnetic field increases, the photoluminescence decreases. At approximately 510 Gauss (see Figure \ref{ESLAC}), there is energy matching between the $P_1$ electronic spin transition and the NV electronic spin transition, facilitating flip-flop interactions that further reduce NV photoluminescence. This specific dip occurs only when the magnetic field is closely aligned with the NV axis. Enhancing the contrast of this dip allows for sub-degree precision in alignment. This technique is detailed in \cite{epstein2005anisotropic}. This secondary alignment tool is crucial for achieving sub-degree precision at higher magnetic fields. As the magnetic field approaches the ESLAC condition, the ODMR contrast for misaligned classes diminishes due to state mixing, making the ODMR technique less effective and more time-consuming.

\section{The influence of $P_1$ centers}

Ideally, optimal polarization of the nuclear spins would occur at the ESLAC. However, at 510 Gauss the energy splitting between $\ket{0_e}$ and $\ket{-1_e}$ in the ground state matches the splitting between $\ket{+1/2}$ and $\ket{-1/2}$ states of the $P_1$ centers. This is shown in Fig. \ref{ESLAC}. 
This matching enables flip-flop processes between $P_1$ and NV centers' electronic spins which drastically reduces both the ODMR contrast and nuclear spin polarization efficiency.
The experiments that are presented in the main text are thus performed under smaller magnetic fields \cite{Jarmola_1}.

\section{Nuclear spin polarization in a static bulk diamond}

 \begin{figure}[h]
\centering
\includegraphics[width=15cm]{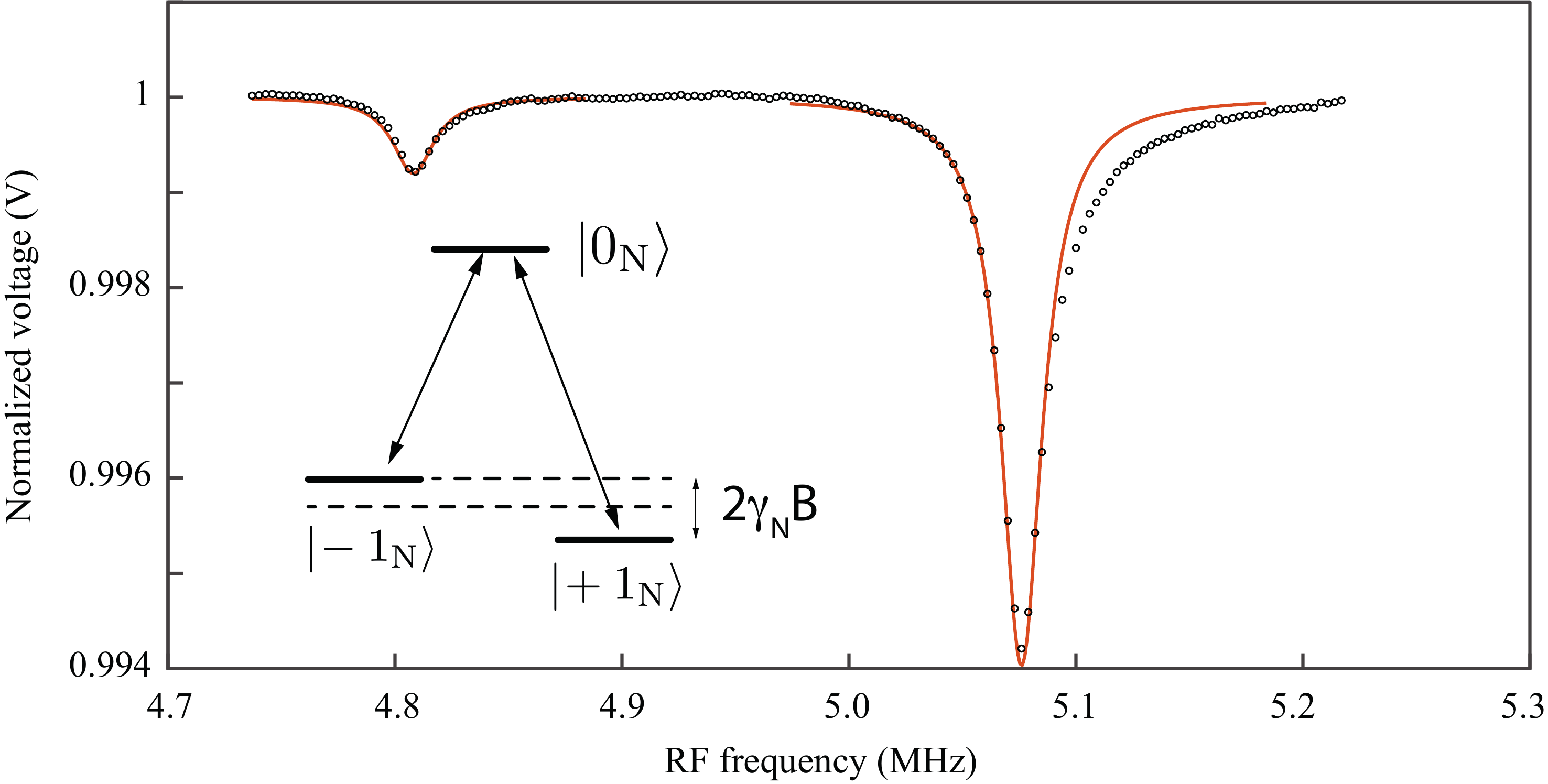}
\caption{Photoluminescence from NV centers in a diamond outside the trap as a function of the RF frequency across the nitrogen nuclear spin resonances. Level structure of the nuclear spin of the $^{14}$N atom indicating the quadrupolar splitting $Q_N=-4.94$ MHz and the Zeeman splitting between the two $\ket{\pm 1_N}$ states.}\label{ODMR_bulk}
\end{figure}

Fig. \ref{ODMR_bulk} shows the photoluminescence of a diamond outside the trap as a function of an RF field across the nitrogen nuclear spin resonances. 
This experiment was carried in a similar way than in the trap, under similar magnetic fields. 
A major difference is the optical power that was used: from 100 $\mu$W in the trap experiment, to 5~mW in the experiment shown in Fig. \ref{ODMR_bulk}.

The two resonances are found to be at  4.808 MHz and 5.076 MHz, indicating an operation at a magnetic field $B=436$~G. The contrast of the transition at 
5.076~MHz is 0.6~\%. This is 30 times larger than the contrast observed under levitation, a discrepancy that can be explained by the lower optical power that was used in the levitation experiment. 
The width of both resonances is about 11 kHz. These are slightly power broadened in order to optimize the sensitivity at the optical power levels we use. 
As in the levitation experiment, the lines were also seen to be asymmetrical. Such an asymmetry was seen to be more and more pronounced as the frequency difference between the two lines decreases, pointing towards simultaneous excitation of both transitions at a given RF frequency.

\section{Rabi and Ramsey oscillations}

Rabi oscillations, as shown in the main text, are important for coupling resonantly to the motion in the dressed state basis, as explained in the main text. 
They however do not provide a precise account of the $T_2^*$ time. Notably, in the main text, the measured decay time of the Rabi oscillations is $T_1^\rho\approx~840 \mu$s. The Rabi decay rate $1/T_1^\rho$ is thus smaller than $1/T_{2,N}^*$. The likely reason for such a small damping of the Rabi oscillations is a spin-locking mechanism that protects the nuclear spins from the fluctuating $P_1$ electronic spin bath when $\Omega_{RF}>1/T_{2,N}^*$.

\begin{figure}[h]
\centering
\includegraphics[width=9cm]{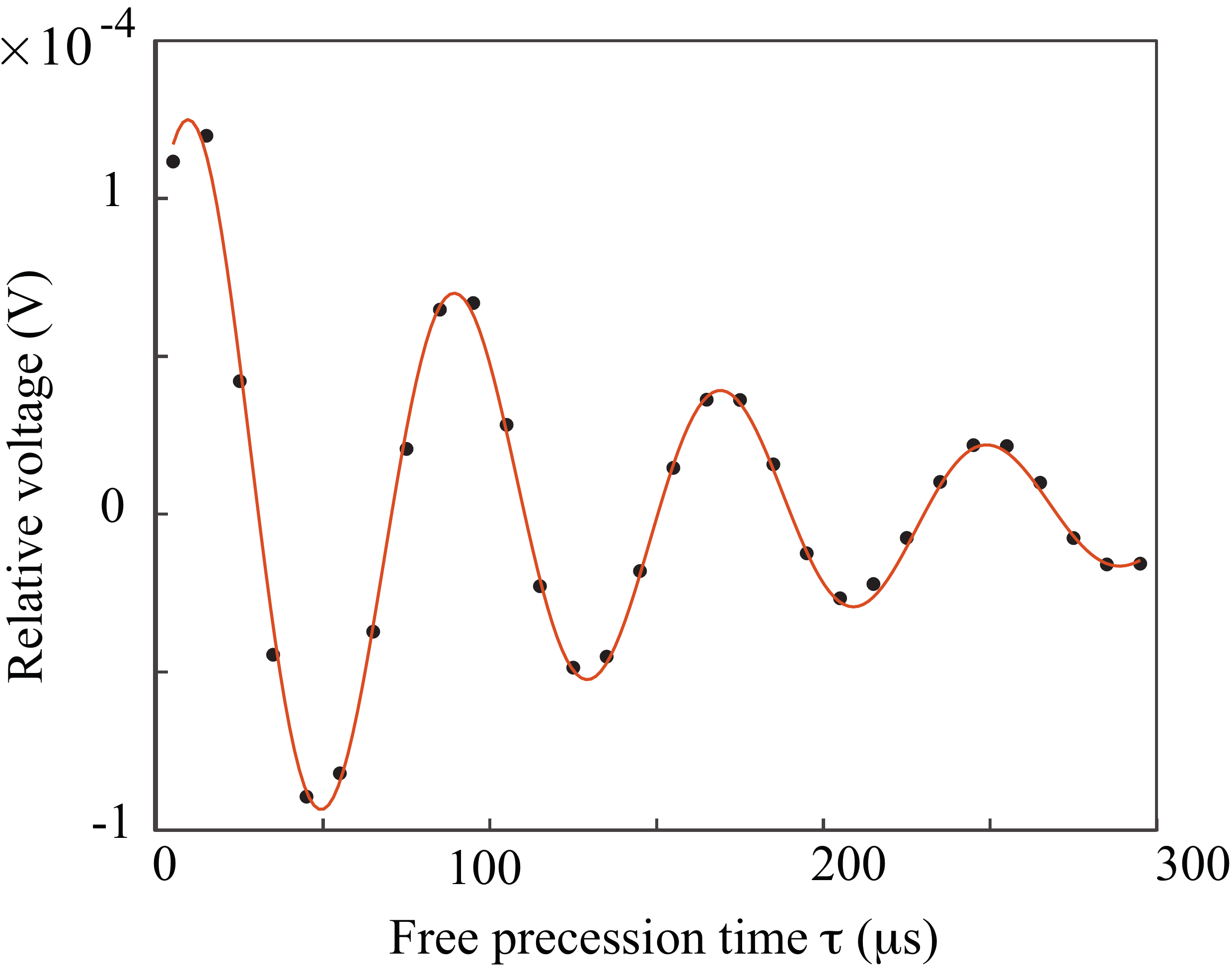}
\caption{Photoluminescence of a diamond outside the trap in a Ramsey sequence. }\label{ramsey_bulk}
\end{figure}

We also performed pulsed measurements on bulk diamonds to compare with the levitation results. 
Spin-locking was also systematically observed when the Rabi frequency was large. 
Fig. \ref{ramsey_bulk} shows the photoluminescence of a diamond outside the trap, as a function of time in the dark in a Ramsey sequence. 
Here the radiofrequency signal was detuned by $\delta=(2\pi) 50$ kHz. 
A decay time of 137~$\mu$s is observed here. 
The larger decay time compared to the experiment presented in the main text can be attributed to the fact that the magnetic field was chosen here to be further away from the $P_1$ lines, which increases slightly the nuclear spin coherence time. Note that in the pulsed measurements, a delay is added after the last microwave pulse in order for the measurement to be taken with the duration (see sequence in the Fig. 3 of the main text). This minimises errors caused by fluctuating powers in the AOM or by the diamond motion, after applying the first laser light pulse.

\bibliography{SI_NMR.bib}